\documentstyle[twocolumn,amssymb,aps,psfig]{revtex}
\tighten
\begin{document}
\draft

\title{\bf Elementary Excitations in Dimerized and Frustrated Heisenberg Chains}

\author{G.~Bouzerar$^1$, A.P.~Kampf$^2$, and G.I.~Japaridze$^3$}

\address{ 
$^1$Institut f\"ur Theoretische Physik, Universit\"at zu K\"oln,\\  
Z\"ulpicher Str. 77, D--50937 K\"oln, Germany \\
$^2$Theoretische Physik III, Elektronische Korrelationen und Magnetismus,\\
Universit\"at Augsburg, D--86135 Augsburg, Germany\\
$^3$ Institute of Physics, Tamarashvili Street 6, Tbilisi 380077, Georgia
 }
\address{~
%\begin{abstract}
\parbox{14cm}{\rm
\medskip
We present a detailed numerical analysis of the low energy excitation 
spectrum of a frustrated and dimerized spin $S=1/2$ Heisenberg chain. In 
particular, we show that in the commensurate spin--Peierls phase the 
ratio of the singlet and triplet excitation gap is a universal function 
which depends on the frustration parameter only. We identify the 
conditions for which a second elementary triplet branch in the excitation 
spectrum splits from the continuum. We compare our results 
with predictions from the continuum limit field theory . 
We discuss the relevance of our data in connection with recent experiments on $CuGeO_{3}$, $NaV_2O_5$, and $(VO)_2P_2O_7$.
\\ \vskip0.05cm \medskip PACS numbers: 71.27.+a, 75.40.Mg, 75.90.+w
}}
%\end{abstract} 
\maketitle

\narrowtext

\section{INTRODUCTION}
Low dimensional quantum spin systems have attracted considerable 
attention of theorists over the decades. Most of the interesting and 
fascinating features observed in these systems are pure quantum effects 
uniquely due to their low dimensionality. Peculiar properties of 
one-dimensional quantum antiferromagnets like e.g. exotic ground states or 
unconventional excitation spectra are not accessible to traditional methods 
like spin-wave or perturbation theory, but require the use of numerical or 
field-theoretical approaches. These methods are complementary to each other 
and together with exact Bethe ansatz solutions of particular models they 
allow for a complete description of low dimensional quantum spin systems. In 
particular, the field-theoretical methods have been used successfully to 
predict the scaling behaviour of the one dimensional spin $S=1/2$ Heisenberg 
model \cite{Affleck1}, the existence of gapless and gapped phases in the 
$S=1/2$ frustrated Heisenberg chain \cite{Haldane1}, and the existence of an 
excitation gap in the spin $S=1$ Heisenberg chain \cite{Haldane2}. On the 
other hand only numerical methods allow to determine the critical value of 
frustration beyond which the gapped phase appears \cite{Okamoto,Castilla} 
and to determine details of the ground state properties \cite{TonegHarada} 
or the behaviour of the excitation gap itself \cite{Chitra,White}.

Recently the interest in one-dimensional spin systems has been 
particularly boosted by the discovery of various non-organic quasi 
one-dimensional compounds, in particular, the spin--Peierls materials 
$CuGeO_3$ \cite{Hase} and $NaV_{2}O_{5}$ \cite{Isobe,Weiden,Regnaultrev,Bray} and spin ladder compounds 
like $SrCu_2O_3$, $Sr_2Cu_3O_5$ or possibly $(VO)_2P_2O_7$ \cite{Garrett1,Garrett2,Luthi,Dagrice}. 
A common feature of these compounds is an excitation spectrum which is 
dramatically different from the spin $S=1/2$ Heisenberg chain. A remarkable 
fact about the Heisenberg chain is that its excitation spectrum consists of 
spin-1/2 particles (spinons) \cite{Takhfadd}. Physically such excitations can 
be created only in pairs because upon flipping one spin the total spin 
projection is changed by $\Delta S_{z} =1$. Thus, in the Heisenberg chain the 
conventional magnons carrying spin $1$ are deconfined into spin-1/2 spinons. 
In dimerized spin-Peierls compounds the excitation spectrum is always gapped and the low lying excitations are triplets.
 In addition, a massive singlet branch may exist above the triplet 
excitation branch in frustrated systems. As it will be shown in 
the following, even a second triplet branch can appear below the continuum. 
Hence in these systems spinons are confined back into triplet magnons.
The interaction between magnons can lead to massive singlet and triplet excitations below the continuum \cite{ShastrySuther,Uhrig} and even a sequence of further massive excitations \cite{Affleck}. 
Furthermore the dimerized frustrated Heisenberg model can also describe the
2-leg ladder with frustration \cite{Brehmer}.

In recent years the field-theoretical continuum-limit approaches were 
successfully used to study spin-Peierls compounds and spin-ladder systems 
\cite{Haldane1,SNT,Tsvelik}. These studies show mechanisms for spinon 
confinement from the alternation of exchange couplings in spin-Peierls 
compounds and from the inter-chain coupling in spin-ladder systems. Although 
universal features of the physical system are usually properly captured in 
field theoretical studies, important details governed by the physics at short 
length scales remain out of range for the applicability of these methods. 
Moreover, due to the perturbative nature of the continuum-limit approach, its 
predictions are less accurate in the physically more realistic strong coupling 
limit where details of the short distance physics are very important. 
Therefore, there is still a number of open questions motivating further 
theoretical studies of spin-Peierls and spin-ladder systems -- especially in 
the strong coupling limit -- by using exact methods.

In this paper we present specifically a detailed numerical analysis of the low 
energy excitation spectrum of the $S=1/2$ antiferromagnetic Heisenberg 
chain with frustration and dimerization, as proposed in particular to describe 
the magnetic properties of $CuGeO_{3}$ \cite{Castilla}. The Hamiltonian reads,
%%%%%%%%%%%%%%%%%%%%%%%%%%%%%%%%%%%%%%%%%%%%%%%%%%%%%%%%%%%%%%%%%%
\begin{equation}
H=J\sum_{i}\left([1+\delta(-1)^i]{\bf S}_i\cdot{\bf S}_{i+1} + 
\alpha{\bf S}_i \cdot {\bf S}_{i+2}\right)
\label{hamilt}
\end{equation}
%%%%%%%%%%%%%%%%%%%%%%%%%%%%%%%%%%%%%%%%%%%%%%%%%%%%%%%%%%%%%%%%%%
where $i$ denotes the sites of a chain with length $L$ and 
${\bf S}_i$ are $S= 1/2 $ spin operators. $J>0$ is the nearest--neighbor 
exchange coupling, $\alpha$ the frustration parameter from next--nearest 
neighbor coupling and $\delta$ is the dimerization parameter. 

Besides its relevance to real spin-Peierls compounds the model is interesting 
purely from a theoretical point of view as far as it contains two independent 
mechanisms for spin-gap formation. At $\delta=0$ the model is characterized 
by a critical value of frustration $\alpha_c$ \cite{Haldane1} which was 
accurately determined by numerical studies: $\alpha_c=0.2412$ 
\cite{Okamoto,Castilla}. For $\alpha < \alpha_c$ the frustration 
is irrelevant, the system renormalizes to the Heisenberg fixed point: the 
ground state corresponds to a spin liquid and the elementary excitations are 
massless spinons. At $\alpha=\alpha_c$ there is a transition into a 
spontaneously dimerized ground state. The spectrum acquires a gap 
and the elementary excitation is a massive spinon \cite{Haldane1}. On the 
other hand at any $\delta \neq 0$ the singlet groundstate of the model is also 
dimerized with a gap in the spin excitation spectrum, but the elementary 
excitation is a magnon \cite{Haldane1,Tsvelik}. 

For the special case $2\alpha+\delta=1$ the groundstate of the spin Hamiltonian
Eq. (\ref{hamilt}) is known exactly to be a product wavefunction of nearest 
neighbour singlet pairs \cite{MajumdarGhosh,ShastrySuther}. This line in the 
$(\alpha,\delta)$ parameter plane separates two distinct regimes: for 
$2\alpha+\delta-1\leq 0$ the dominant peak in the static magnetic structure 
factor is at $q^{*}= \pi$, while in the other case $\pi/2<q^{*}<\pi$. In this 
latter incommensurate phase, $q^{*}$ continuously decreases from $\pi$ with 
increasing $\alpha$ and $\delta$, and asymptotically approaches $\pi/2$ 
\cite{Chitra,White}. 

In a recent work \cite{paper1}, the existence of a massive singlet excitation 
has been confirmed numerically for the Hamiltonian Eq. (\ref{hamilt}) with 
$\alpha=0.35$ and $\delta=0.012$. In addition to the elementary 
triplet and singlet excitations and depending on the set of parameters 
($\alpha$,$\delta$) even another triplet excitation was found to split from 
the continuum \cite{paper1,Yokoyama}.

 In this paper we will analyze in detail 
the intriguing structure of the excitation spectrum for different frustration 
and dimerization parameters, and we compare our numerical data with the 
available results from field theoretical methods. The paper is organized as 
follows: In chapter II we summarize the procedure and the results of the 
continuum limit field theory and outline the open questions inaccessible by 
these analytical methods. In chapter III we focus our attention on the 
singlet to triplet energy gap ratio, $R(\alpha,\delta)=\Delta_s/\Delta_t$ and 
show that $R(\alpha,\delta)$ only depends on $\alpha$ when $2\alpha+\delta<1$; 
the field theory prediction $R=\sqrt{3}$ is precisely realized for all 
$\delta$ {\it only} when $\alpha=\alpha_{c}$. In chapter IV we discuss the 
conditions for which a second triplet excitation branch may exist below the 
continuum. Finally, in chapter V we connect our results to recent experimental 
data on different spin chain compounds.

\section{THE CONTINUUM-LIMIT FIELD-THEORY APPROACH}

First insight into the structure of the excitation spectrum of the model 
Eq. (\ref{hamilt}) is obtained from bosonization and the continuum limit 
renormalization group approach \cite{AfflekHaldane}. 
In terms of the continuum field $\phi(x)$ the bosonized version of the initial 
spin model is the double sine--Gordon (SG) model 
%%%%%%%%%%%%%%%%%%%%%%%%%%%%%%%%%%%%%%%%%%%%%%%%%%%%%%%%%%%%%%%%%%
\begin{eqnarray}
H_{bos}=\int{\rm d}x\, \Big[\frac{u}{2}&&[K\, \Pi^{2} + 
K^{-1}\left(\frac{d\phi}{dx}\right)^{2}] \nonumber \\
&+& M_{\delta} \cos(\beta_{\delta}\phi) + M_{\alpha} \cos(\beta_{\alpha}\phi)\Big]
\label{hamiltbos}
\end{eqnarray}
%%%%%%%%%%%%%%%%%%%%%%%%%%%%%%%%%%%%%%%%%%%%%%%%%%%%%%%%%%%%%%%%%%
where u is the spin wave velocity, $\beta_{\alpha}=\sqrt{8\pi}$,
$\beta_{\delta}=\sqrt{2\pi}$,
$K=\sqrt{\frac{1-(\alpha-\alpha_{c})}{1+(\alpha-\alpha_{c})}}$,
$M_{\alpha} = J(\alpha-\alpha_{c})$, and $M_{\delta} = \delta$ . 

The value of the critical frustration $\alpha_{c}$ is determined by 
the behaviour of the system at short distances i.e. it depends on nonuniversal 
parameters of the continuum limit theory. Therefore, differently constructed 
continuum limit theories give rather different values of this parameter 
\cite{Haldane1,Nakano,Bishop}. The exact value of $\alpha_{c}$ was determined 
only within numerical studies \cite{Okamoto,Castilla}, but the non-universal 
parameters of the continuum limit Hamiltonian could be always chosen in such 
a way to ensure the proper value of the critical frustration $\alpha_{c}$. 

Contrary to the standard SG model (Hamiltonian \ref{hamiltbos} with only 1 
'cosine term': $M \cos(\beta \phi)$), which is exactly solvable and 
well understood \cite{Sklyaninetc}, the theory of the quantum double 
SG model is much less developed. However, in two limiting cases the 
model Eq. (\ref{hamiltbos}) reduces to the SG theory and provides 
exact knowledge about the characteristic properties of the system.

Let us first consider the case $\delta = 0$  ($M_{\delta}=0$)
corresponding to the frustrated 
Heisenberg chain.
The behaviour of this model is determined by the 
marginal interaction which is controlled by the frustration:$\beta=\beta_{\alpha}=\sqrt{8\pi}$. For $\alpha < 
\alpha_{c}$ the interaction is irrelevant and the system scales to the 
Gaussian fixed point: elementary excitations are massless spinons. For 
$\alpha > \alpha_{c}$ the interaction is marginally relevant and the effective 
interaction renormalizes to large values. An exponentially small gap 
$M^{*}\propto\exp(-c/(\alpha-\alpha_{c}))$ is dynamically generated in the 
excitation spectrum, the field $\phi$ is ordered leading to a spontaneously 
dimerized ground state with the finite order parameter 
%%%%%%%%%%%%%%%%%%%%%%%%%%%%%%%%%%%%%%%%%%%%%%%%%%%%%%%%%%%%%%%%%%
\begin{equation}
\langle\hat{O}_{d}\rangle={1\over L}\langle\sum_{i}(-1)^{i}
({\bf S}_{i-1}\cdot{\bf S}_{i}-{\bf S}_{i}\cdot{\bf S}_{i+1})\rangle\,\, . 
\label{Ordpar}
\end{equation}
%%%%%%%%%%%%%%%%%%%%%%%%%%%%%%%%%%%%%%%%%%%%%%%%%%%%%%%%%%%%%%%%%%
Thus, the elementary excitations in the massive phase of the 
frustrated Heisenberg chain are described by solitons ('kinks') of the 
quantum SG model with $\beta=\sqrt{8\pi}$ \cite{Haldane1}. There are no 
soliton-antisoliton bound states in this case and the system is characterized 
by the only one scale - spin gap $\Delta = 2M^{*}$. 
Excitations above the given vacuum are created by breaking singlet bonds. 
Each broken bond gives rise to a pair of decoupled spins $1/2$ on 
neighbouring sites. Once created, these isolated spins can propagate 
coherently along different sublattices and constitute elementary excitations
of the massive spinon type.

We now consider the case $\delta\neq 0$ and $\alpha=\alpha_{c}$
($M_{\alpha}=0$ and $M_{\delta} \neq 0$ )
The excitation spectrum of 
the SG model at $\beta = \beta_{\delta}=\sqrt{2\pi}$ is exactly known 
\cite{Sklyaninetc} and at this point consists of soliton and antisoliton 
excitations with masses $M_{s}=M_{\bar{s}}=M$ and two bound states 
(breathers) with masses $M_{1}=M$ and $M_{2}=\sqrt{3}M$. The soliton 
excitation carries spin $S^{z}=1$, the antisoliton excitation $S^{z}=-1$, 
and the two breathers with opposite parity $S^{z}=0$. The lower energy breather
mode is degenerate with the kink and anti-kink and these three excitations 
correspond to a 
triplet excitation branch in the original spin model language. The second 
bound state, in fact, has its counterpart in a spin singlet excitation 
\cite{Haldane1}. These two modes are the only elementary excitations in this 
case and the ratio of their excitation gaps is exactly $\sqrt{3}$. 

The standard Renormalization Group (RG) approach 
\cite{Tsvelik,Bishop,Giamarchi}
to the double SG model Eq. (\ref{hamiltbos}) is based on the fact that the 
critical dimensions of the two cosine terms arising from the smooth and 
staggered part of the exchanges, respectively, are different:
%%%%%%%%%%%%%%%%%%%%%%%%%%%%%%%%%%%%%%%%%%%%%%%%%%%%%%%%%%%%%%%%%%
\begin{equation}
{\rm dim}\, \cos(\sqrt{2\pi}\phi) = 1\hskip0.2cm,\hskip0.2cm
{\rm dim}\, \cos(\sqrt{8\pi}\phi) = 2\,\, .
\label{Dimensions}
\end{equation}
%%%%%%%%%%%%%%%%%%%%%%%%%%%%%%%%%%%%%%%%%%%%%%%%%%%%%%%%%%%%%%%%%%
Thus, the $\delta\cos(\sqrt{2\pi}\phi)$ term is strongly relevant, while the 
$J(\alpha-\alpha_{c})\cos(\sqrt{8\pi}\phi)$ term is marginal. Therefore, the 
essential physics as determined by the relevant term is -- at least for 
$\delta\ll 1$ -- similar to that of the above discussed SG model with 
$\beta=\sqrt{2\pi}$, and the marginal interaction leads to logarithmic 
corrections only. Therefore one assumes that the excitation spectrum of the 
spin-Peierls state consists of two excitation branches with gaps $\Delta_{t}$ 
(triplet excitation) and $\Delta_{s}= R \Delta_{t}$ (singlet excitation) 
with $R$ slightly different from $\sqrt{3}$ due to the logarithmic corrections
\cite{Tsvelik,Giamarchi}. However, since frustration and dimerization provide two 
principally different mechanisms for spin gap formation, interference between 
these interactions is non-trivial especially in the limit of strong initial 
interactions. A very sensitive tool to study these particular effects is to 
explore the detailed structure of the excitation spectrum. Moreover, the exact 
excitation spectrum of the dimerized and frustrated Heisenberg chain Eq. 
(\ref{hamilt}) provides another way for the determination of the critical 
parameter $\alpha_{c}$. Only for $\alpha=\alpha_{c}$ is the structure of the 
excitation spectrum exactly the same as that of the SG model with 
$\beta=\sqrt{2\pi}$. Therefore, $\alpha_c$ is determined from the condition 
$R(\delta,\alpha_{c})=\sqrt{3}$. 
 
Due to the different critical dimensions of the cosine terms in the double SG 
model one may attempt, in a first approximation, to neglect the 
$\cos(\sqrt{8\pi}\phi)$ term and to consider the usual SG model with the 
$\cos(\sqrt{2\pi}\phi)$ term only. However, as we demonstrate in the subsequent
chapters, we have to conclude from our numerical results that this commonly 
accepted procedure is not valid and the structure of the excitations is quite 
different. E.g. we find $R=2$ in the absence of frustration $\alpha=0$ which   
means that there is no long wavelength singlet excitation branch and the 
singlet excitation energy coincides with the edge of the continuum. Furthermore
we obtain that for $\alpha<\alpha_{c}$ and small $\delta$ the ratio 
$R=\Delta_s/\Delta_t$ is a 'universal' function of $\alpha$ alone. So although 
the continuum limit Hamiltonian is a proper description of the spin lattice 
model Eq. (\ref{hamilt}) the commonly adopted field theoretical tools for the 
double SG model are not sufficient for a complete understanding of the 
excitation spectrum as we will show from our exact diagonalization data.

\section{GAP RATIO}
Our numerical study is performed using exact diagonalization techniques with 
periodic boundary conditions for chains with up to $L=26$ sites. As previously 
reported in Ref. \cite{paper1} the triplet and singlet gaps $\Delta_t$ and 
$\Delta_s$ have different finite size scaling behaviour. While $\Delta_t$ is a 
monotonically decreasing function of $1/L$, $\Delta_s$ is non--monotonic in 
$1/L$ and develops a minimum for a particular chain length which varies with 
the model parameters. Thus, in order to extrapolate the values for $\Delta_t$ 
and $\Delta_s$ to the infinite chain limit, we need two different finite size 
scaling fit functions. For the triplet gap we have used the 3--parameter 
ansatz \cite{paper1},
\begin{equation}
\Delta_t(L)=\Delta_t+\frac{A}{L}\exp{\left(-\frac{L}{L_t}\right)}
\,\, .
\label{fiteq1}
\end{equation}
On the other hand, in order to account for the non--monotonic behaviour of 
$\Delta_s$ we have chosen the 4--parameter ansatz
\begin{equation}
\Delta_s(L)=\Delta_s+\exp{\left(-\frac{L}{L_s}\right)}(\frac{A}{L}+B)
\label{fiteq2}
\end{equation}
with $A>0$ and $B<0$. Note that both fit functions proved to give an excellent 
agreement with density matrix renormalization group (DMRG) data with $L$ of 
order 100, in particular in the region where the spin-spin correlation length 
is shorter than the chain length \cite{paper1,Schonfeld}.

In Fig.\ref{fig1} we show the ratio $R(\alpha,\delta)$ as a function of 
$\alpha$ for different values of the dimerization parameter 
($0.02<\delta<0.4$) using the extrapolated values of $\Delta_t$ and $\Delta_s$.
Since we find that the width of the singlet dispersion is smaller than the 
width of the triplet dispersion, the necessary condition for the singlet 
excitation branch to split from the continuum over the whole Brillouin zone is 
$R(\alpha,\delta)<2$. Fig.\ref{fig1} clearly shows that the ratio depends on 
the frustration parameter in an {\it essential} way. For $\alpha<\alpha_{c}$ 
we obtain $R(\alpha,\delta)>\sqrt{3}$ implying that the $\cos(\sqrt{8\pi}\phi)$
term in the double SG model is indeed relevant for all $\alpha\ne\alpha_c$ and 
can not be omitted. We recall the field theoretical expectation that the  
deviations from $\sqrt{3}$ should be logarithmically small. Furthermore, the
exact diagonalization data show that the ratio $R$ is for $\alpha<\alpha_c$ 
insensitive to the dimerization parameter $\delta$ (to be more precise, for 
$\delta>0.05$ within the accuracy of our finite size studies) which in the 
continuum limit field theory controls the relevant interaction in the double SG
model. $R=\sqrt{3}$ is indeed obtained at criticality $\alpha=\alpha_c$ for 
any finite $\delta$, weak or strong. In the absence of frustration, $\alpha=0$,
and for any dimerization we find $R(\alpha=0,\delta)=2$. Therefore, there is 
no well defined singlet excitation at $q=0$ (or equivalently $q=\pi$) since
$\Delta_s=2\Delta_t$ coincides with the lower edge of the continuum. 

In addition we find that (i) $R(\alpha,\delta)$ is a {\it universal} function
of $\alpha$ for $2\alpha+\delta\leq 1$ and (ii) $R(\alpha,\delta)$ depends
on both parameters in the incommensurate phase for $2\alpha+\delta>1$. We note 
that the deviations observed for weak values of $\delta$ (i.e. $\delta=0.05$) 
in Fig.\ref{fig1} result from a lack of precision in the determination of the 
extrapolated $\Delta_s$. High precision is lost when the spin--spin correlation
length becomes comparable to or longer than the chain length for small 
$\delta$. On the other hand the comparison with DMRG calculations for 
$L\leq 100$ shows that the ansatz for the scaling of the singlet gap 
$\Delta_s$, Eq. (\ref{fiteq1}), remains very accurate even in the region of 
long correlation lengths (small gap). To demonstrate the consistency with 
the diagonalization data we have added in Fig.\ref{fig1} DMRG data points calculated 
for $\delta=0.02$ \cite{Schonfeld}. The DMRG data support the observation that
the ratio $R(\alpha,\delta)$ depends on $\alpha$ {\it only} in the commensurate
phase, i.e. as long as $2\alpha+\delta\leq 1$. We have indicated in Fig.\ref{fig1} 
(with a cross) the data points for which the parameter pairs $(\alpha,\delta)$ 
belong to the incommensurate phase. We emphasize that for these sets of 
parameters the finite size effects are extremely small and the values of 
$R(\alpha,\delta)$ are thus very accurate. 

A similar behaviour has recently been found by Yo\-ko\-yama et al. for the 
leading $\delta$ power law dependence of the triplet excitation gap 
$\Delta_t\propto\delta^{\gamma}$, where the exponent $\gamma$ is a monotonic 
continuous function of $\alpha$ \cite{Yokoyama}. In this work it was shown that
the Cross-Fisher value $\gamma=\frac{2}{3}$ \cite{Cross} is realized only for 
$\alpha=\alpha_{c}$, but $\gamma\neq\frac{2}{3}$ for $\alpha<\alpha_{c}$.

\section{A SECOND TRIPLET BRANCH}
Another peculiar observation is made when the parameters $(\alpha,\delta)$ are 
increased towards stronger dimerization in that a second triplet excitation 
branch splits from the continuum. In analogy to the singlet excitation this 
second triplet may be interpreted as a bound-state between a triplet and a 
singlet excitation. In order to investigate this new feature we fix the 
dimerization parameter $\delta=0.2$ and discuss the low energy spectrum as a 
function of frustration in the subsector of total momentum $q=0$. For this 
purpose we have show in Fig.\ref{fig2}a the energies of the 3 lowest excited 
states, extrapolated to $L\rightarrow\infty$, as a function of $\alpha$. In 
this figure the lower edge of the continuum at $2\Delta_t$ is indicated by a 
continuous bold line. We observe that for $\alpha<0.28$, there are only two 
well defined excitations below the continuum, one triplet and one singlet as 
discussed above. The energy of the next excited state is found to scale to 
the lower edge of the continuum with $L\rightarrow\infty$. In contrast to the 
triplet, the singlet excitation gap remains almost constant with changing 
$\alpha$. However, we observe that for $\alpha>0.28$ another triplet excitation
$T_{2}$ splits from the continuum (i.e. $E_{T_{2}}<2\Delta_t$). This is in 
agreement with the results reported previously in Ref. \cite{paper1}.

In order to verify that the triplet $T_2$ is indeed a well defined elementary 
excitation at $q=\pi$ only when $\alpha>0.28$ (for $\delta=0.2$), we have also
evaluated its spectral weight $W=|\langle T_2|\hat{O}|0\rangle|^2$ versus $1/L$
in the dynamical structure factor $S(\pi,\omega)$ for different frustration 
parameters $\alpha$ (see Fig.\ref{fig2}b) where $|0\rangle$ is the groundstate
wave function and $\hat{O}=S^z(\pi)=(1/\sqrt{L})\sum_l\exp{({\rm i}\pi\,l)}
S^{z}_l$. We observe that for $\alpha<0.28$, $W$ has a strong size dependence, 
and the data indicate that $W\rightarrow 0$ with $L\rightarrow\infty$. However,
for $\alpha>0.28$ the curvature of $W(1/L)$ changes indicating that the weight 
$W(1/L)$ scales to a finite value consistent with the identification of $T_2$ 
as an elementary excitation below the continuum.

By evaluating the excitation spectrum in different momentum sectors we find 
that the triplet $T_2$ splits from the continuum first at $q=\frac{\pi}{2}$. In
Fig.\ref{fig3}a, we have plotted the ratio,
\begin{equation}
R_{1}=\frac{E(S=1,\frac{\pi}{2},2)-E(S=1,\frac{\pi}{2},1)}{\Delta_t}
\label{r1}
\end{equation}
versus $\alpha$ where $E(S,q,n)$ is the $n$--th energy level in the subsector 
with total momentum $q$ and spin $S$ extrapolated to $L\rightarrow\infty$.  
$R_{1}< 1$ implies that the excitation is split from the continuum at 
$q=\frac{\pi}{2}$. Fig.\ref{fig3}a shows that for large enough dimerization 
(e.g. $\delta=0.2$) the triplet $T_{2}$ is well defined at $q=\frac{\pi}{2}$ 
for any $\alpha$. However, as we reduce the dimerization parameter (e.g. to 
$\delta=0.1$), we observe that stronger frustration is needed in order to 
separate $T_2$ from the continuum. In fact, for $\delta=0.1$ and $\alpha<0.12$,
$R_1\approx 1$, i.e. the energy of the second triplet excitation scales to the 
lower edge of the continuum.  
Thus our data suggest that in the absence of 
frustration, there is a minimal value $\delta_{min}>0.1$ above which the 
triplet $T_2$ appears. The fact that $R_1 \approx 0.98$ instead of 1, for $\delta=0.1$ and $\alpha< 0.12$ shows that this excitation is not well defined, the finite size analysis is not valid for excitations in the continuum.
The kink in $R_1$ clearly indicates the minimum value of the frustration 
for which the excitation separates from the lower edge of the continuum.
Similarly, the value $\alpha=0.28$ (see Fig. \ref{fig2}a) is the 
minimal frustration $\alpha_{min}(\delta=0.2)$ for the 
second triplet to be a well defined excitation branch over the entire 
Brillouin zone. In other words, for a given $\delta$, the momentum range for 
which the triplet $T_2$ is split from the continuum is centered around 
$q=\pi/2$ increasing continuously with increasing $\alpha$. As an example, we 
show in Fig.\ref{fig4} the dispersion of the 3 lowest excitations for 
$\alpha=0.2$ and $\delta=0.2$, for which the triplet $T_2$ is split from the 
continuum in the momentum range $4\pi/5>q>q_{min}\approx\pi/5$.

In order to determine the $\delta$-dependence of $\alpha_{min}$, we have 
plotted in Fig.\ref{fig3}b the ratio
\begin{equation}
R_{2}=\frac{E(S=1,\pi,2)-E_{0}}{\Delta_t}
\label{r2}
\end{equation}
versus $\alpha$ for different values of $\delta$. The excitation $T_2$ is 
split from the continuum when $R_{2}<2$. Fig.\ref{fig3}b shows that 
$\alpha_{min}$ increases when $\delta$ is reduced, for instance 
$\alpha_{min}(0.2)\approx 0.28$ and $\alpha_{min}(0.05)\approx 0.32$.
We have plotted in  Fig.\ref{fig3}c  $\alpha_{min}$ as a function of $\delta$
for $ \delta < 1$. Our results suggest that $\alpha_{min}$ is within a good approximation a simple linear function of the parameter $\delta$, 
$\alpha_{min} \approx  (1-\delta)/3$.
Thus, in the special limiting case $\delta \rightarrow 1$ , and restricting ourself to the commensurate region we observe that $\alpha_{min} \rightarrow 0$.
Note that the result can not be extended at the special point $\delta=1$ , since the spectrum at this point consists of a discrete set 
of eigenvalues $E_n=n*\Delta_{t}$, thus there is no continuum in the vicinity of $2*\Delta_{t}$. However, for $ 0.5 < \delta < 1$, the spectrum consist of mini-bands with finite band-gaps, thus the notion of continuum has a meaning in the vicinity of $2*\Delta_{t}$. Furthermore our result suggests that the spectrum is completely dense (no band gap), when $ \delta < 0.5$.
The other interesting point corresponds to the limit $\delta \rightarrow 0$. Indeed, if the linear approximation is still valid in this limit, our data shows that $\alpha_{min} \rightarrow 1/3$.

\section{Discussion of experimental data.}

We now try to connect these results to some recent experimental data on 
$CuGeO_{3}$, $NaV_{2}O_{5}$ and $(VO)_{2}P_{2}O_{7}$. $CuGeO_{3}$ \cite{Hase} 
and $NaV_{2}O_{5}$ \cite{Isobe,Fujii,Weiden} are spin-Peierls systems with 
transition temperatures $T_{SP}\approx 14K$ and $35K$, respectively. Previously
it has been shown that the magnetic susceptibility of $CuGeO_3$ in the uniform
phase can be accurately reproduced by a frustrated Heisenberg chain model with 
\cite{Fabricius}; from this fit the frustration was estimated to be 
$\alpha\approx 0.35$, close to the previous estimate of Riera and Dobry 
\cite{Riera}. With an exchange coupling $J\approx 160K$ and $\alpha=0.36$ the 
experimental value for the gap $\Delta_t=2.1 meV$ as determined by inelastic 
neutron scattering \cite{NishiRegnault} is obtained within the frustrated and 
dimerized Heisenberg chain model with a dimerization $\delta=0.012$ 
\cite{paper1}. With this parameter set a singlet--triplet gap ratio 
$R(0.35,0.012)\approx 1.50$ follows (see Fig.\ref{fig1}).
However, the experimental ratio is 
$R_{exp}=1.72\approx\sqrt{3}$; according to Fig.\ref{fig1}, this would 
corresponds to $\alpha\approx\alpha_{c}$, a value for $\alpha$ which was 
previously proposed by Castilla et al. \cite{Castilla}.

There are different possibilities how to resolve this quantitative problem.
First, the effects of interchain coupling along the crystal $b$-direction in
$CuGeO_3$ are experimentally well established but little is known from 
theoretical studies \cite{coupled-chains}. Second, the dynamics resulting from the 
spin-lattice coupling is expected to play an important role. Indeed, 
including dynamical phonons will renormalize the spin excitation 
spectrum and as a direct consequence the ratio R \cite{AugierPoilblanc}.

Recently, it has been proposed that the low temperature phase of the 
compound $NaV_{2}O_{5}$ is well described by an unfrustrated ($\alpha=0$) 
dimerized chain model \cite{Isobe}. Following the same procedure as used for 
$CuGeO_{3}$, the dimerization was estimated to be of order 
$\delta\approx 0.048$ \cite{Augier}. If we assume that this is indeed the case
then -- according to our results -- we expect a singlet-triplet gap ratio 
$R(0,0.048)=2$, i.e. there is no well defined long wavelength singlet 
excitation below the continuum. As a consequence, we expect no magnetic Raman 
response at frequencies below the lower edge of the continuum at $2\Delta_t$.
However, in a recent Raman scattering experiment on $NaV_{2}O_{5}$ 
\cite{Lemmens} it has been observed that the energy of the lowest excitation 
in the dimerized phase is at $64 cm^{-1}$; for this compound the measured 
triplet excitation gap from inelastic neutron scattering is 
$\Delta_t\approx 59cm^{-1}$ \cite{Weiden}. Thus the energy of the lowest 
excitation in the Raman spectra is very close to the singlet-triplet gap 
$\Delta_t$. Furthermore, it was observed that this feature remains unchanged 
for different photon polarization geometries suggesting that this excitation 
is probably not of magnetic origin. An alternative origin is a charge 
excitation due to a local charge transfer from a $V^{4+}$ to a $V^{5+}$ chain.
All these observations indicate some important qualitative differences between 
the spin chain compounds $CuGeO_{3}$ and $NaV_{2}O_{5}$.

Finally we discuss the case of $(VO)_{2}P_{2}O_{7}$ -- a spin chain compound 
whose magnetic properties have given rise to controversial interpretations. 
This compound has been initially considered as an ideal realization of a 
two-leg antiferromagnetic Heisenberg spin ladder \cite{Johnson}. Subsequently, 
however, it has been shown that the susceptibility data on polycristalline 
material could be well fitted by either a ladder or an alternating chain model 
\cite{Barnes}. Early neutron scattering data indicated a spin gap of about 
$50K$ and supported the two-leg ladder picture \cite{Eccleston}. Recently, 
inelastic neutron scattering experiments have been performed on a collection 
of many oriented single crystals \cite{Garrett2}. The observed features like 
e.g. the strong dispersion in the rung direction lead to the conclusion that 
the data are not consistent with a spin ladder description and that 
$(VO)_{2}P_{2}O_{7}$ is better described as an alternating spin chain directed 
in the rung direction. Remarkably, in addition to the lowest triplet another 
well defined higher energy triplet excitation below the continuum was observed 
near $q=\pi$. If we assume that this compound is well described by an 
unfrustrated ($\alpha=0$) dimerized chain with a dimerization of order 
$\delta\approx 0.1$ (i.e. an alternating exchange coupling in the rung 
direction), then we find a contradiction with our present results. In fact, 
for $\delta\approx 0.1$ and $\alpha=0$, there is no well defined second 
triplet at $q=\pi$. The microscopic structure of $(VO)_2P_2O_7$, however, does
not allow to identify any obvious superexchange path which can lead to 
frustration. At this point we can only point out that both pictures -- a 
two-leg ladder or an unfrustrated alternating chain -- do not properly 
describe the low energy excitation spectrum in this compound. The correct
modeling of the magnetic properties of $(VO)_{2}P_{2}O_{7}$ remains an 
unresolved problem.

\section{Summary and Conclusion}

In this paper we have studied the elementary excitations of frustrated and 
dimerized Heisenberg chains. We have shown that in the commensurate dimerized 
phase the singlet-triplet gap ratio $R(\alpha,\delta)$ only depends on the 
frustration parameter for arbitrary values of the dimerization 
($\delta\ge 0.02$). The magnitude of this ratio gives a direct measure for the 
strength of the frustrating exchange coupling. Without frustration 
$R(\alpha=0,\delta)=2$ implying the absence of a long wavelength singlet 
excitation below the continuum. We have shown that away from criticality 
$\alpha\neq\alpha_{c}$ the frustration term plays an important role for the 
low energy excitation spectrum. Therefore, the commonly adopted procedure of 
disregarding the frustration term in the bosonized continuum limit Hamiltonian 
for $\alpha<\alpha_{c}$ is at least questionable since it is not marginally 
irrelevant. However, for $\alpha=\alpha_{c}$, our exact diagonalization results
give precisely $R(\alpha_{c})=\sqrt{3}$ in perfect agreement with the field
theoretical treatment of the SG model. Furthermore, we have demonstrated the 
conditions for the existence of another triplet excitation branch which we 
interpret as a bound state of elementary triplet and singlet excitations. 
Depending on the dimerization strength the second triplet excitation branch is 
observable at $q=0$ (or $q=\pi$) only if the frustration parameter is large 
enough. As we have discussed we consider our results relevant for the magnetic
excitations in the spin chain compounds $CuGeO_3$, $NaV_2O_5$, and 
$(VO)_2P_2O_7$. For the latter material we conclude from our results that 
neither a two-leg ladder nor an alternating, unfrustrated Heisenberg chain
model captures the experimental facts on the excitation spectrum in 
$(VO)_2P_2O_7$.

\centerline{{\bf Acknowledgements}}
We gratefully acknowledge helpful discussions with B. B\"uchner, H. Fehske, D. 
Khomskii, A. Kl\"umper, P. Lemmens, E. M\"uller-Hartmann, D. Poilblanc
 and S. Sil. We especially thank F. Sch\"onfeld for providing us with unpublished DMRG data.
This research was performed within the program of the Sonderforschungsbereich 
341 supported by the Deutsche Forschungsgemeinschaft.

%
%  Fig. 1
%

\begin{figure}
\caption[]{
Extrapolated singlet-triplet gap ratio $R(\alpha,\delta)=$ $\Delta_s/\Delta_t$ 
vs. frustration $\alpha$ for different dimerization parameters $\delta$.
The symbols marked with a cross have been calculated in the incommensurate 
phase. ED indicates exact diagonalization results extrapolated to the 
infinite chain limit. In addition we have added data points from density
matrix renormalization group (DMRG) calculations \cite{Schonfeld}.
}
\label{fig1}
\end{figure} 

%
%  Fig. 2
%

\begin{figure}
\caption[]{
(a) Energy of the 3 lowest excitations vs. frustration calculated at fixed 
dimerization $\delta=0.2$ and momentum $q=0$ (resp. $\pi$). The continuous 
line indicates the edge of the continuum at $2\Delta_t$. $T_{1,2}$ are triplet 
excitations and $S_{1}$ is the lowest singlet excitation.
(b) Spectral weight W of the triplet $T_{2}$ as a function of $1/L$ (L is the size of the system) for $\delta=0.2$ and different values of $\alpha$.
The full symbols correspond to cases for which W scales to finite values
in the infinite chain limit.

}
\label{fig2}
\end{figure} 

%
%  Fig. 3
%

\begin{figure}
\caption[]{
(a) $R_{1}$ and (b) $R_2$ as a function of $\alpha$ calculated for different 
values of the parameter $\delta$. For the definitions of the ratios $R_1$ and 
$R_2$ see Eqs. (\ref{r1}) and (\ref{r2}), respectively.
In fig (c), we have plotted $\alpha_{min}$ as a function of $\delta$, the dashed line is a linear fit of the data.

}
\label{fig3}
\end{figure} 

%
%  Fig. 4
%

\begin{figure}
\caption[]{
Dispersion $\omega_{i}(q)= E_{i}(q)-E_{0}$ of the three lowest excitations 
for $\alpha=0.2$ and $\delta=0.2$. The full symbols correspond to the triplets 
$T_{1}$ and $T_{2}$ and the open symbols to the singlet $S_{1}$. The continuum 
is indicated by the shaded area. 
}
\label{fig4}
\end{figure}

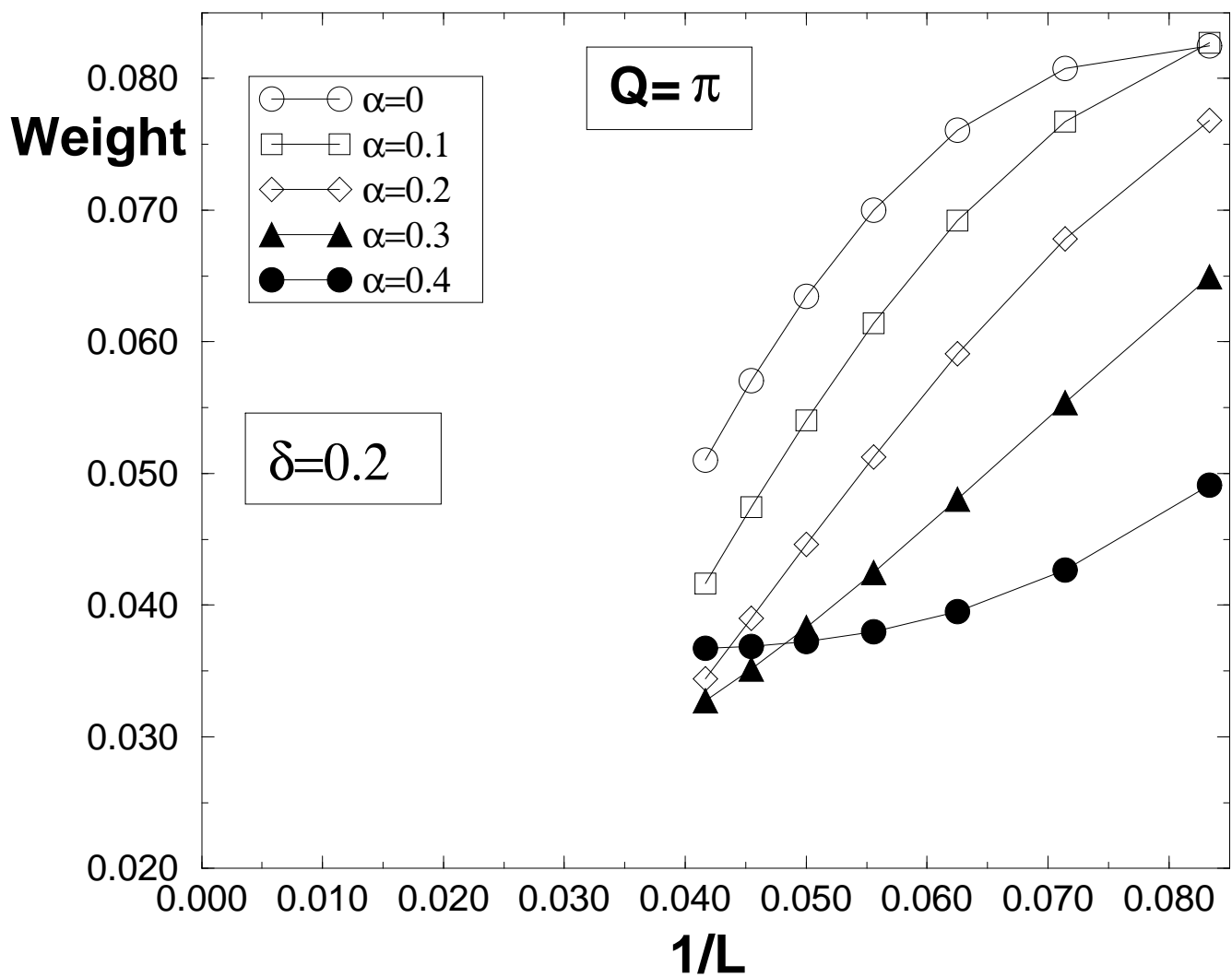
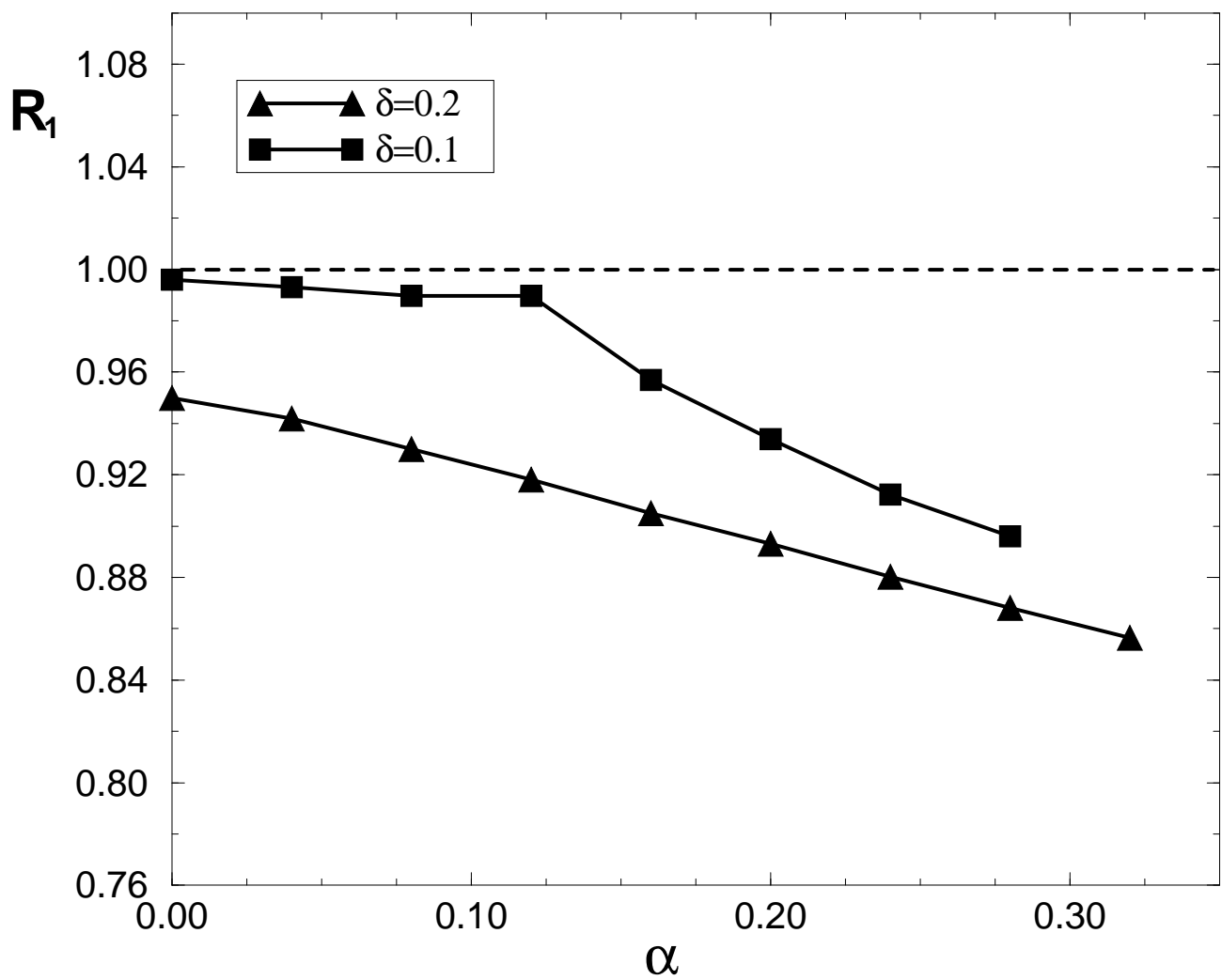
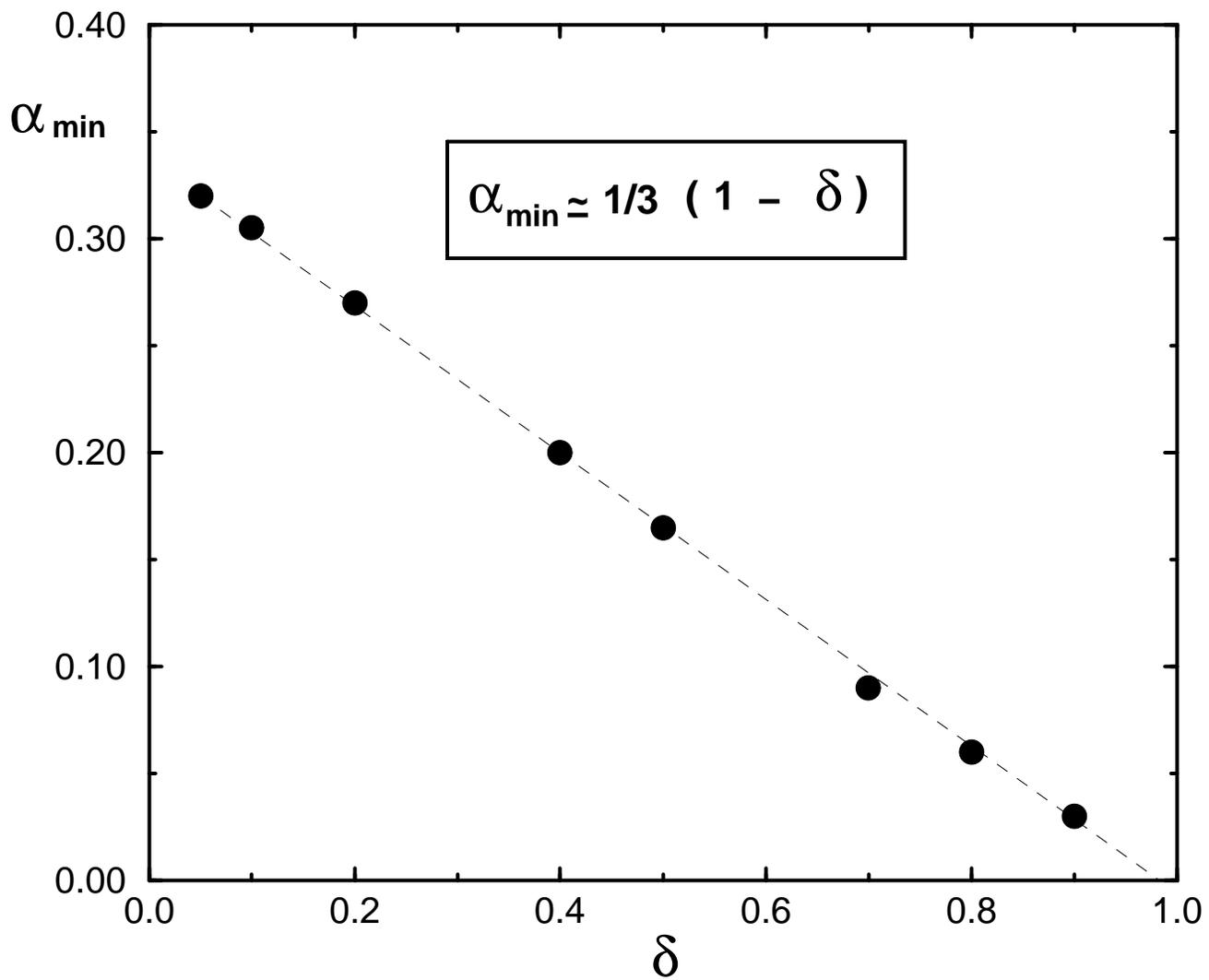
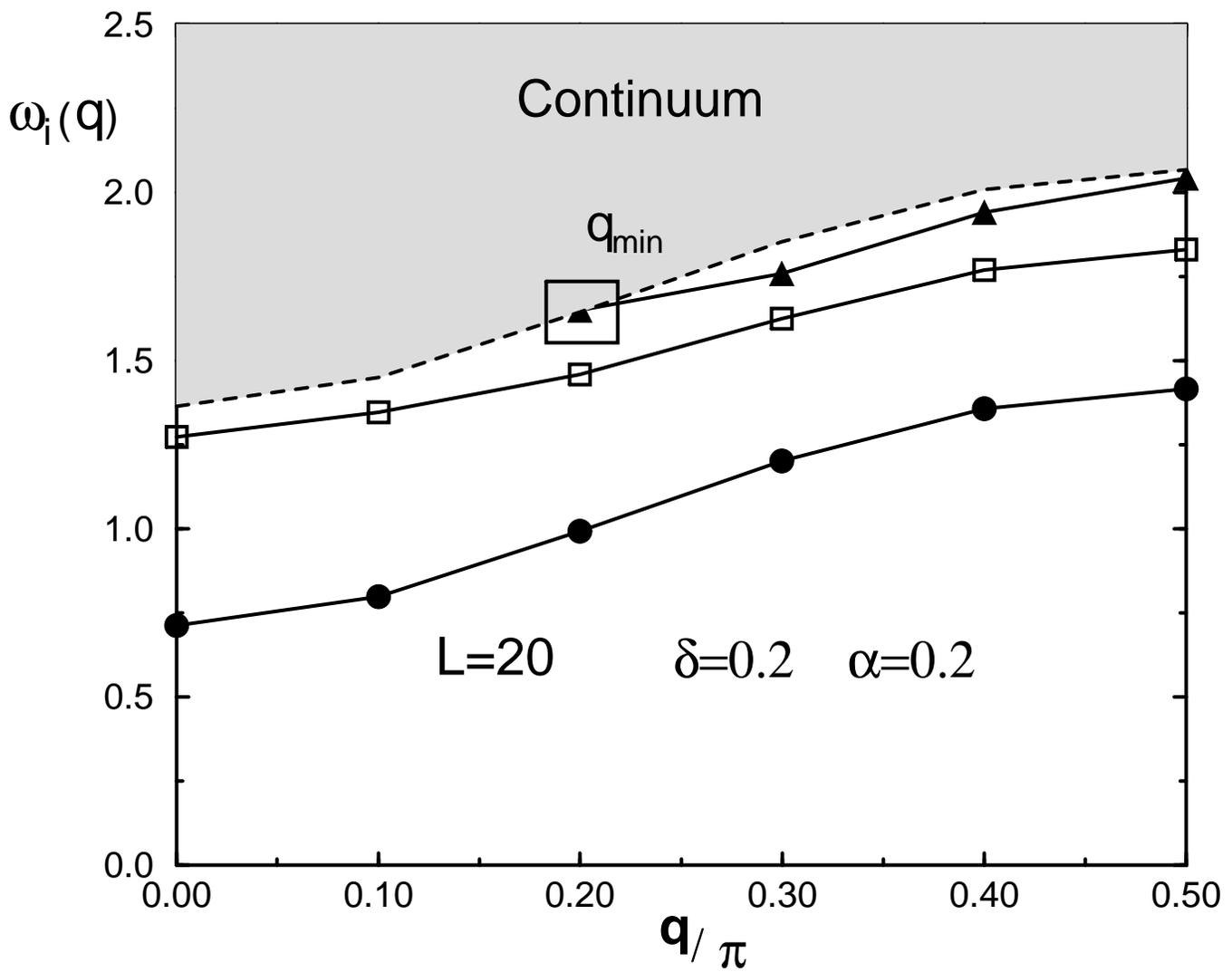
\end{document}